\documentclass[preprint,5p,times]{elsarticle}

\usepackage{amsmath}

\usepackage{amssymb}

\journal{Future Generation Computer Systems}

\begin{document}

\begin{frontmatter}



\title{Mahiru: a federated, policy-driven data processing and exchange system}


\author[inst1]{Lourens E. Veen}

\affiliation[inst1]{organization={Netherlands eScience Center},
            addressline={Science Park 402}, 
            city={Amsterdam},
            postcode={1098XH}, 
            country={The Netherlands}}

\author[inst2]{Sara Shakeri}
\author[inst2]{Paola Grosso}

\affiliation[inst2]{organization={Multiscale Networked Systems, University of Amsterdam},
            addressline={Science Park 904}, 
            city={Amsterdam},
            postcode={1098XH}, 
            country={The Netherlands}}

\begin{abstract}
Secure, privacy-preserving sharing of scientific or business data is currently a popular topic for research and development, both in academia and outside of it. Systems have been proposed for sharing individual facts about individuals and sharing entire data sets, for sharing data through trusted third parties, for obfuscating sensitive data by anonymisation and homomorphic encryption, for distributed processing as in federated machine learning and secure multiparty computation, and for trading data access or ownership. However, these systems typically support only one of these solutions, while organisations often have a variety of data and use cases for which different solutions are appropriate. If a single system could be built that is flexible enough to support a variety of solutions, then administration would be greatly simplified and attack surfaces reduced. In this paper we present Mahiru, a design for a data exchange and processing system in which owners of data and software fully control their assets, users may submit a wide variety of processing requests including most of the above applications, and all parties collaborate to execute those requests in a distributed fashion, while ensuring that the policies are adhered to at all times. This is achieved through a federated, mostly decentralised architecture and a powerful policy mechanism designed to be easy to understand and simple to implement. We have created a proof-of-concept implementation of the system which is openly available and in continuous development, and which we aim to continue to extend with new functionality.
\end{abstract}

\begin{graphicalabstract}
\includegraphics[width=\textwidth]{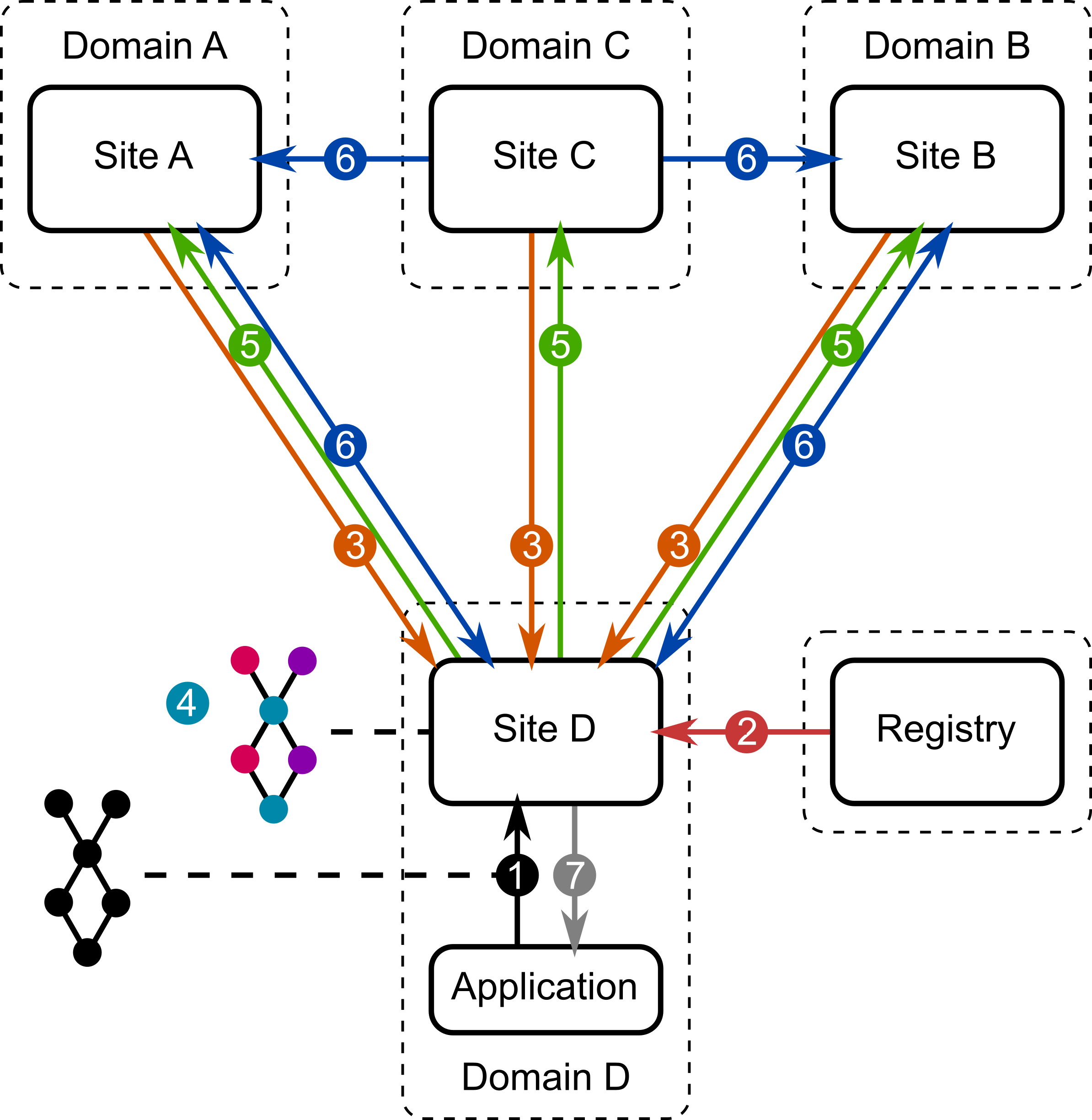}
\end{graphicalabstract}

\begin{highlights}
\item Federated infrastructure for data exchange and privacy-preserving data analysis
\item Novel policy mechanism for federated workflow execution
\end{highlights}

\begin{keyword}
data sharing \sep federation \sep policies \sep workflows
\end{keyword}

\end{frontmatter}


\section{Introduction}
\label{sec:introduction}
Every day, vast amounts of data are processed in business and in science to support operations or, through \emph{data science}, to gain understanding of the world behind the data. Recently, interest has developed in obtaining new insights by combining data across different organisations. In some cases, simple data exchange between parties can fulfill this requirement, but often the data involved are privacy- or commercially sensitive, and the parties are legally unable or strategically unwilling to provide others with a copy. In such cases, collaboration can still occur, but more advanced techniques are needed.

A variety of such techniques have been proposed, all of them involving some form of distributed computation. One approach is to somehow obfuscate sensitive data before sharing it, through anonymisation, (homomorphic) encryption or secure multiparty computation. Potential downsides to these methods are imperfect protection and poor performance. Another option is to not share the data at all, but allow processing to be done wherever the data is and share only aggregated results. Compute-to-data solutions and federated machine learning algorithms fall in this category. Trusted third party systems combine remote processing with simple sharing, allowing processing of the raw data to take place at some location not managed by either the data provider or the data consumer.

Various systems have been and are currently being designed to address each of these solutions. However, they typically only support one of these models, so that an organisation wishing to do all of them, perhaps across different data sets and with different partners, will end up having to maintain a variety of systems. Furthermore, existing systems often require a significant amount of trust between the participating parties in order for the system to work at all, with data access treated in an all-or-nothing fashion. As a result, organisations may end up running multiple copies of the system in order to serve different collaborative projects.

Mahiru is a design for a federated, policy-driven data processing and exchange system for use by data scientists. It runs generic workflows, and as such supports all the distributed data processing patterns mentioned above as well as arbitrary combinations of them. Owners of data and software set policies which govern whether and where their data or software can be copied and how data can be processed. These are enforced by their own system and the systems they choose to share copies with. Data scientists use data and software by running local applications which submit workflows to their local Mahiru site, which plans them and executes them in collaboration with the other sites in the system. Throughout the execution, policies are enforced by all the involved sites.

This paper is organised as follows. Section \ref{sec:related_work} describes related work, covering a selection of existing data sharing systems as well as addressing the concept of trust in data sharing. Section \ref{sec:design_considerations} sets out the key design considerations that informed the design of Mahiru. In section \ref{sec:architecture} the overall architecture of Mahiru is described as well as how it operates. Section \ref{sec:policies} describes the policy mechanism in general, with Section \ref{sec:use_cases} providing examples of how it can be used. In Section \ref{sec:implementation} we discuss the implementation in some detail, including some networking and security considerations. Finally, in Section \ref{sec:discussion}, we discuss the strengths and weaknesses of the design and present some possibilities for future development.

\section{Related work}
\label{sec:related_work}

\subsection{Data sharing systems}
\label{sec:data_sharing_systems}
A great many projects are currently underway to create different kinds of cross-organisational data sharing systems. These systems vary widely in scope and architecture. Rather than trying to review all of them, we will describe a few representative systems that illustrate the different goals and approaches taken. A first distinction that can be made here is between sharing non-sensitive (business) data and sharing sensitive (personal) data. In the former case, data is usually shared by providing others with a copy after negotiating terms and securing payment, while in the latter case (remote) data processing comes into play. Systems also differ in the type of data they share, the common types being individual records (transaction-oriented systems), entire data sets (analytics-oriented systems), or streams (typically Internet-of-Things applications). From a technical perspective, there are differences in the degree of centralisation of data and metadata, and whether there is support for data processing.

Perhaps the largest project in this space is International Data Spaces (IDS), a project initiated by the European Union with the goal of connecting and more closely integrating European industry, in particular SMEs\cite{IDS}. IDS defines data exchange protocols, and provides central components including a data broker, clearing house, identity provider, app store, and vocabulary provider \cite{IDSArch}. Data are requested from a data provider, optionally processed, and returned to the data consumer. Several IDS-based or IDS-supporting data exchange systems are coming online. One of these is the Smart Connected Supplier Network, a system for exchanging order-related data between organisations in a supply chain\cite{SCSN}. It provides a standardised API used to send orders, invoices and despatch information. Participants connect to a provider, which routes their messages in a peer-to-peer fashion to the provider of the recipient, who forwards them again.

A second IDS-compatible project is the Mobility Data Marketplace (MDM), which aims to be the German national access point for traffic data\cite{MDM}. MDM comprises a central portal, through which metadata are made available for streams of data. Parties negotiate for access separately, then permissions are set by the provider and data is streamed through a central data broker. Finally Advaneo is a data marketplace for open and commercial data, with a centralised metadata store, a payment system using credit cards, centralised data exchange for open data and peer-to-peer data exchange through IDS connectors for more sensitive commercial data.

The Ocean Protocol takes a very different approach, and focuses on trading ownership of or access rights to data\cite{ocean_protocol}. It stores metadata on a blockchain, and enables trade on the blockchain in tokens representing data or data access rights, in a manner similar to a commodities market. Obtaining the data corresponding to a token is mostly outside the scope of the project, which focuses on market making and derivatives more than data access or processing.

The above systems are mostly geared towards trading commercial or public data between companies and governments. When it comes to sharing of personal data, an oft-mentioned project is SOLID\cite{SOLID}. SOLID gives individuals a secure digital locker, in which they can store information about themselves. These data can then be copied by other parties from the locker, but only if the person concerned permits them to do so. In contrast to the above systems, SOLID deals in individual records rather than data sets or streams. There is no payment system built in and data processing is done by the recipient once it has a copy of the record(s). Other, similar digital locker systems are also being developed, both for personal and for business data\cite{Dexes}\cite{genetic_locker}.

A final category of systems for sharing personal data do distributed data analysis. Data Shield is an R-based data analysis system which allows a user to interactively access one or more remote data sets and calculate summary statistics, but only in such a way that no data about individuals is revealed\cite{data_shield}. It does this by offering only a limited API, and by using an implementation that protects against differential privacy violations. Users need individual accounts on the data servers they access, and authenticate using certificates. Vantage6 is a federated machine learning system implementing the Personal Health Train concept\cite{Vantage6}. It assumes a collaboration between a number of organisations that wish to share data. Each data provider sets up a data station, which gets access to one or more data sets. A central server orchestrates a federated machine learning algorithm, repeatedly sending partially trained models to the data stations for further training, until the model coefficients converge. The trained model is then returned to the user, who can use it to make predictions without having had access to the data. Parties within a collaboration are assumed to trust each other.

\subsection{Trust}
\label{sec:trust}
A successful data sharing system is a system that data owners and data users are willing and able to use. Convincing data owners to make their data available is usually the largest hurdle to clear. The design of a data sharing system should therefore be informed by an understanding of how a data owner decides whether to share their data. Data sharing is a social interaction, which can only happen in the presence of sufficient trust.

Trust can be defined as \textit{the willingness of a party to be vulnerable to actions of another party based on the expectation that the other party will perform a particular action important to the trustor, irrespective of the ability to monitor or control that other party}\cite{mayer_1995}. Without the ability to monitor or control the other party, trusting them inevitably incurs risk. Conversely, one can say that any social interaction involving a risk that cannot be fully mitigated requires trust. When deciding whether to trust or not, potential trustors will evaluate the associated risks and look for \textit{good reasons} to accept those risks, i.e. factors reducing either the (perceived) probability or the consequences of an adverse outcome. If enough good reasons can be found that increase trust and decrease risk until the level of available trust exceeds the remaining risk, the decision to trust is taken\cite{bachmann_2001}.

Thus, for a data sharing system to be used, it needs to offer features that increase the users' trust and decrease risk. Trust can be divided into two categories: system trust and personal trust. System trust or institutional trust is created by legal or social norms, and backed up by the power of institutions including the judiciary, trade associations and technical standards. Note that laws or codes of conduct do not actually make adverse outcomes impossible, and that violating them will destroy trust in that specific instance, but as long as they are generally adhered to and violations are sanctioned in general, they create trust in and throughout the system.

Personal trust concerns the specific counterparty in a social interaction. A potential trustor may attempt to evaluate the specific trustworthiness of the trustee by considering their benevolence, integrity and ability\cite{mayer_1995}. \textit{Benevolence} is the degree to which the trustor believes the trustee to want to do right by them, separate from any reward the trustee may get for doing so. This requires the trustee to have the same understanding of what is right, which is referred to as \textit{integrity}. Finally, a benevolent potential trustee of high integrity needs to have the \textit{ability} to execute on their good intentions to be trustworthy.

In an increasingly complex world with many interactions, personal trust tends to be replaced by system trust because personal trust is very expensive to maintain. If system trust is unavailable, organisations may instead choose to base their interactions more on power than on trust. In an interaction based on trust, a positive scenario is constructed with the expectation that both parties will work to enact it. In an interaction based on power, a negative scenario is constructed and the more powerful party uses the threat of sanctions to keep the subordinate party from enacting it. This is obviously a less desirable situation for the subordinate party, and may lead them to refuse the interaction. Thus, in the context of data sharing systems, strong system trust, backed up by institutions, encourages less powerful parties to take part and share their data\cite{van_der_burg_2021}.

\section{Design considerations}
\label{sec:design_considerations}
Mahiru is designed for doing data science, both in academia and in industry, using data that is not necessarily public but shared between parties conditionally. Its intended users are researchers, and providers of data, software, services and infrastructure. To enable their collaboration, Mahiru needs to provide the necessary technical functionality while reducing risk and increasing trust as much as possible.

\subsection{Risk}
\label{sec:design_considerations_risk}

The fundamental risk of giving someone else a copy of your data is in \textit{loss of control}. This loss of control can then result in loss of privacy and loss of economic opportunity (if the copy is shared further without permission), as well as reputation damage and exposure to sanctions (if one is accountable for misuse to some third party). There are fundamentally two technical means to mitigate this risk: 1) try to assert some level of control over the receiving party's system, and 2) don't give them a copy at all but instead allow them to process the data on the system of the data owner or that of a trusted third party.

For option 1), remote attestation and other Digital Restrictions Management (DRM) technology would be used to ensure that the receiver's system runs trusted software. These technologies cannot attest however that the hardware and software that are used to store and process the data do not have any vulnerabilities, so while DRM can reduce risk, it cannot remove it. In practice, DRM technologies have not proven very effective at preventing unauthorised use of widely distributed data, as only one receiver needs to be able to circumvent the technology and then publish the content, but they may help for data that is shared only with a small number of known users. They are however a system administration burden on the data receiver, and they reduce the receiver's autonomy in administrating their system.

A fundamental design choice of Mahiru is that participants in the data exchange should retain full control over their systems. Accordingly, DRM technologies are not required, and Mahiru's security model does not depend on their presence. Nevertheless, Mahiru users could choose to implement e.g. remote attestation separately, and they may find others more willing to share data with them if they do so.

Option 2), distributed processing, only applies to cases where the data user actually needs data that is derived from the source data, and the derived data is less sensitive than the source data. Also, the data user now needs to trust the data provider (or a third party) with their processing code, and the data provider needs to trust the processing code to output only the intended less sensitive output. The associated risks may be mitigated by inspecting, sandboxing and/or monitoring the processing code, or by inspecting the output before sending it back to the receiver. An advantage here is that the data remains on the system of its owner, and under their full control.

In order to support option 2), the system needs to be able to execute processing workflows in a distributed fashion, and it needs an authorisation system which can describe where data can go and how it can be processed. This significantly increases system complexity, but also allows fundamentally better privacy protection for many use cases. Mahiru provides both distributed execution and an authorisation system for it, as described below.

\subsection{Trust}
\label{sec:design_considerations_trust}

Besides mitigating risk, measures can be taken to increase trust. Perhaps the most important one is setting up behavioural norms that participants in the data exchange are expected to abide by. International Data Spaces is the prime example of this, having invested from an early stage in what they call \textit{soft infrastructure}. The power needed to back up the norm can be provided by legal contracts signed by participants as a condition for joining the data exchange, and for example by having an external party audit their systems from time to time. At the technical level, this means that it must be possible to limit access to the exchange to parties that have followed an appropriate process and have passed inspection. Mahiru supports this by letting a single party control its registry of participating parties and sites (see below), but the actual terms and conditions are outside its scope.

Trust in software applications can be increased by inspecting and certifying them, increasing specific trust in the certified applications. Using Mahiru, auditors that are part of an exchange can mark software providers' software as trustworthy, and data owners can then give permission to process their data based on those marks (see Section \ref{sec:delegation}). Additionally, system trust can be increased by monitoring execution of the applications as they run on other parties' systems, through the prospect of anomalous behaviour being detected and sanctioned. We discuss this further in Section \ref{sec:implementation}.

To help facilitate personal trust, the system can provide participants with information about who controls the data, software and systems that make up the data exchange. This transparency helps participants to evaluate benevolence, integrity and ability of those parties. To this end, Mahiru stores and exchanges metadata about parties, sites, data and software. Some kind of reputation tracking system could be added in order to facilitate exchange of relevant information between the parties, possibly enhancing trust although such a system could also be attacked by malicious parties. This is currently outside our scope.

No discussion of data sharing and trust can be complete without mentioning blockchain technology. Blockchains are said to be \textit{trustless}, in the sense that they do not require any participant to (personally) trust any other specific participant. Perhaps it is better to say that blockchain technology provides a large amount of system trust, backed by technical standards, a consensus algorithm, and a community of participants, which reduces or even removes the need for personal trust. However, it is important to understand that this trust only pertains to the integrity and evolution of the contents of the shared database formed by the blockchain. Thus, a blockchain may record that two parties have expressed a desire to share data under certain conditions, but it cannot enforce those conditions once the data is on the system of the receiver, nor can it for example ascertain that the data has arrived on the system of the receiver. As a result, the use of blockchains in data sharing systems is limited to payment for data (as implemented by Datapace) and trading of ownership rights (as in the Ocean Protocol).

\section{Architecture}
\label{sec:architecture}

Mahiru is designed to be installed once by an organisation, and then used over time to service multiple collaborations. With partners and partners-of-partners joining the system, there will be participants which do not know or trust each other at all, as well as participants which collaborate closely and have a high degree of trust between them. Mahiru takes the no-trust situation as a starting point: it is a federated system in which each participant operates its own site, consisting of its own software, running on its own hardware, sitting on its own premises. The organisation's data and software are stored in this site, and are not available to anyone by default.

Any of these aspects may be relaxed: the servers may be put in a data center operated by someone else, they can be virtual machines running on cloud hardware, software may be obtained from elsewhere and managed by someone else (e.g. Mahiru sites could be offered as a Software-as-a-Service solution), and the Mahiru policy system may be used to give permission for data and software to be transmitted to others or processed on their behalf. None of these are required however, full control is always possible.

A Mahiru data exchange system thus consists of many \emph{sites}, operated by different \emph{parties}. These sites communicate over the Internet, and need to be able to find each other and set up secure communications. This is facilitated by the \emph{Registry}. The registry contains a list of all parties and sites involved in the data exchange, but no data, software or policies. In most cases, it will be a simple database maintained by an organisation running the data exchange, which would verify any general requirements made by the exchange on the parties and their sites (e.g. minimum security measures) in the physical world, and only add those who qualify to the database. Alternatively, anyone could be allowed to register, in which case it would be up to individual parties to decide which other parties and sites to trust.

\subsection{Registry}
\label{sec:registry}

The registry stores information about the parties and sites involved in the data sharing system. Each of these objects needs to be uniquely identifiable. Mahiru does not have a global unique identifier service. Instead, each party uses a DNS name it controls as its namespace, and then creates unique identifiers of the form \texttt{<type>:<ns>:<name>}, where \texttt{<ns>} is its namespace and \texttt{<type>} is either \texttt{party} or \texttt{site}. The \texttt{<name>} part must be unique within the namespace. Note that parties are identified by an identifier within their namespace, not by the namespace itself. Furthermore, departments or subsidiaries can be registered as separate parties with their own namespace, and grouped under their parent party, as described in Section~\ref{sec:organising_objects} below.

Besides its identifier and namespace, a party registration contains several X.509 certificates. The party's main certificate is used to sign registry records and policy rules. Its User CA certificate is used to sign certificates for individual users, which are also stored in the registry within the party's record. Sites have a unique identifier, and store the identifiers of their owner and administrator. The owner of a site is the party using it, the administrator the party controlling the servers and software. These can be the same, for a self-hosted site, or different if the site is hosted by another party. Finally, each site has an endpoint URL where it can be reached, and an HTTPS certificate for communicating.

If the CA and the Registry are kept separate, which they should, then this provides some level of security redundancy. If the Registry is compromised, records can be removed to achieve a denial-of-service attack, but records cannot be modified without the private key of the subject of the record (which is stored at their site and potentially offline), and no new records can be added without separately compromising the CA to create the corresponding certificates. Likewise, if the CA is compromised, the attacker would still need to compromise the Registry as well for the rest of the system to accept its newly minted fake certificates.

In order to keep the Registry from becoming a central bottleneck in exchanges with very many parties and sites, the information in the Registry is replicated to each site rather than requested every time it is needed. Transactions on the Registry are serialised and numbered, and each object is tagged with the transaction number in which it was created when it is created. Objects are never removed. Instead a deletion operation adds a second tag to the object with the transaction number of the removal, thus marking it as having been removed in that and subsequent versions. Updates are processed as removal and simultaneous addition of the updated object.

Sites then request updates from the current version of their replica to the current version of the canonical store, and receive a set of objects to add, a set of objects to remove, the new version, and a timestamp specifying until when their newly updated replica will be valid. Whenever the local replica is expired, it must first be updated before its contents can be used again.

Instead of a central database and a replication system, a completely decentralised data exchange could be achieved by putting the registry on a (public or private) block chain. Note that no trading or data processing would take place on this block chain, it would only register the parties and sites, and contain a mechanism for admitting new parties to the system. The data exchange would then become a decentralised autonomous organisation.

\subsection{Sites}
\label{sec:sites}

Data storage, exchange and processing is done exclusively by the sites. Each site stores data and software (together, \emph{assets}) as well as the policies governing their exchange, processing and use. Sites offer an external REST API served over HTTPS covering three areas: 1) policy replication, 2) asset sharing, and 3) data processing. Authentication is done through standard HTTPS X.509 certificates provided by an external Certificate Authority; additionally they may be stored in (and removed from and thus invalidated) the Registry and/or OCSP may be employed to facilitate rapid retraction of compromised certificates. Sites also have an internal REST API, which is used for managing assets, managing policies, and submitting data processing requests.

An asset may be copied (downloaded) from one site to another, if its owner has given permission to do so. The receiver of the asset is expected to treat the asset according to its owner's policies. Parties may (partially) delegate policy decisions regarding one or more of their assets to other parties in the system, who can then control the asset by changing their own policies. Thus, in order to know what permissions one has regarding a certain asset, policies from several sources have to be evaluated together. To facilitate this, policies are replicated between the sites using the same replication algorithm used by the Registry. Each site maintains a local replica of all the parties' policies, then evaluates them locally whenever it needs to make a policy decision.

Note that this means that each site's copy of the combined policies is only eventually consistent with the global state of the policies, because policy updates at a given site do not propagate immediately to all other sites. As a result, access may be granted based on recently retracted policies. However, as policy freshness requirements are set by the originating site, each participant can decide for itself whether to set a long timeout (slow propagation, but low load on the replication service) or a short one (fast propagation, higher load and cost).

\subsection{Operation}
\label{sec:operation}

\begin{figure}
\includegraphics[width=\columnwidth]{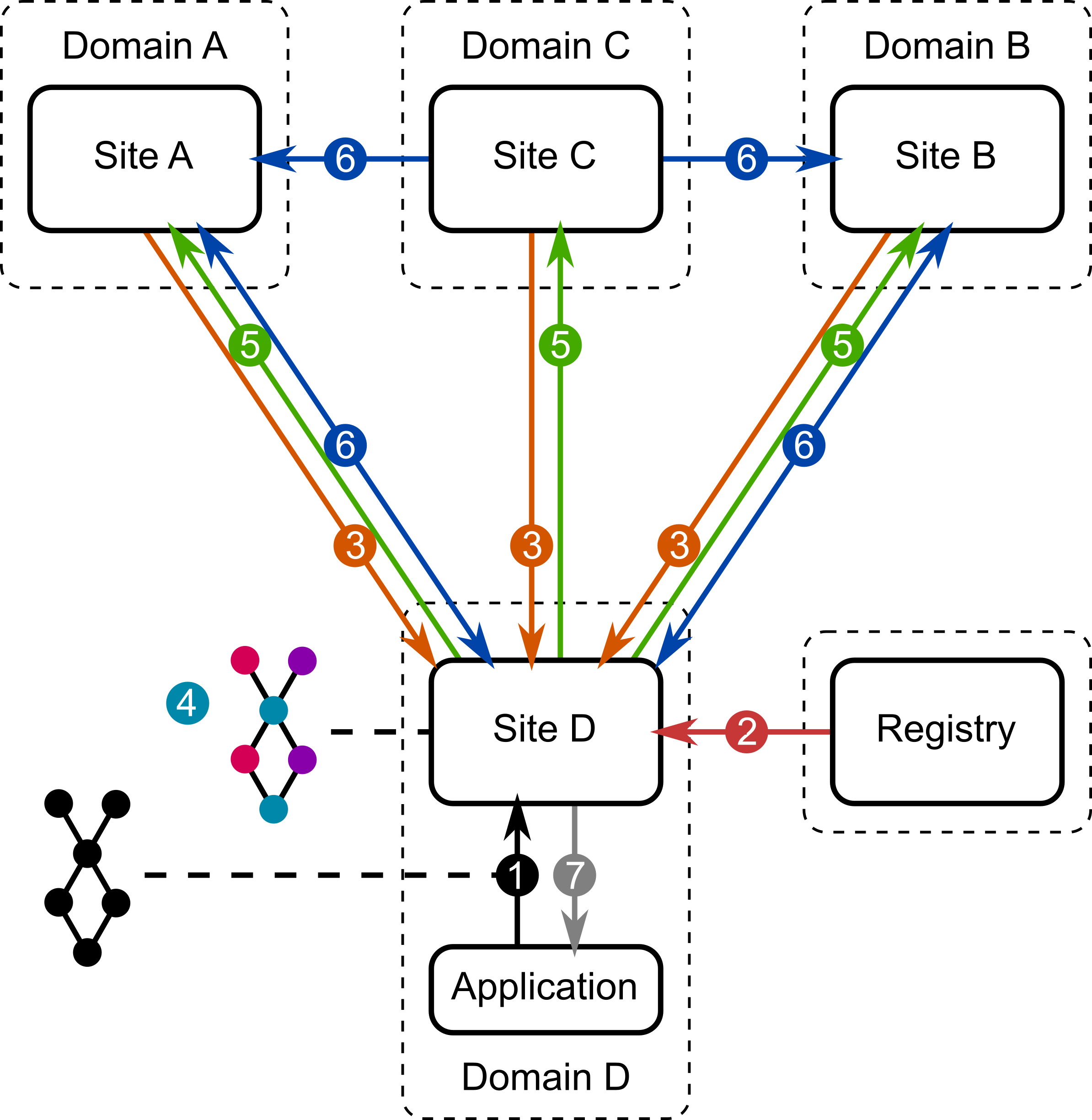}
\caption{Mahiru's global architecture and operation. The system consists of any number of sites and a registry which records their identity and location. 1) Workflow submission, 2) registry update, 3) policy update, 4) planning, 5) execution request, 6) distributed execution, 7) result return. See Section~\ref{sec:operation} for details.}
\label{fig:global_operation}
\end{figure}

Requests for (processing of) data in Mahiru take the form of workflows of the Directed Acyclic Graph (DAG) type, with data flowing along the edges; the format can be described as a simplified version of the Common Workflow Language\cite{CWL}. Each node (workflow step) in the DAG employs a \emph{compute asset} to process one or more \emph{data assets}, producing one or more new data assets which are propagated to subsequent steps. We envision a user using any of a variety of applications with Mahiru support, depending on their context and use case, with the workflow generated by the application rather than by the user directly.

Figure~\ref{fig:global_operation} shows Mahiru's global operation. Execution is initiated by the user's application, which submits a workflow to the local Mahiru site (1). The site then ensures that its Registry replica is valid (2), and that the policy replicas for the sites serving policies for the assets used in the workflow are likewise current (3). It will then create an execution plan (4), which for each step in the workflow determines at which site it will be executed (see Section \ref{sec:policy_evaluation}). Each of the involved sites is then sent an execution request (5), comprising the workflow and the plan. These sites will in turn ensure they have up-to-date policy replicas (not shown), verify that the actions they are asked to perform are permitted, and start executing steps as their inputs become available. During this process, intermediate results may be exchanged between the sites via the asset sharing part of their API (6), each transaction again being checked against the policies set by the data owners. The originating site finally obtains the final outputs from the sites that produced them, verifies that usage permissions exist, and returns the outputs to the user's application (7).

\subsection{Identity management}

Mahiru's users are individuals associated with a party that is involved in the system. System administrators and data managers can be identified and authenticated at each local site using off-the-shelf solutions. Workflow submission however requires a separate solution, since workflow execution is distributed across sites. Running a workflow requires resources, which may have to be paid for, and this requires a means for each party involved in the execution to prove that the workflow was submitted by a given party. Parties in turn may want to know which of their users has submitted which workflow for accounting and accountability purposes, but this information should not be shared with other parties to avoid violating the privacy of its users (employees).

Non-repudiation can be solved by a digital signature on the workflow, which in Mahiru is created by the user using a pseudonymous user certificate and key. This certificate is signed by a signing certificate controlled by the party's system administrator, which in turn is signed by the central CA, with both the party CA and user certificates registered in the Registry. As a result, any party involved in execution can verify the signature, and prove that it was created by a user belonging to a particular party, but it cannot identify the user beyond the pseudonymous identity. This enables accounting and reporting abuse, while protecting employee privacy.

\section{Policies}
\label{sec:policies}

In a Mahiru system, each participating party can contribute assets (software and data) by putting them into its local site. In order to allow others to make use of these assets, policies must be set. Mahiru has a bespoke policy system that occupies a niche in between system-level policies (e.g. file access, firewall rules) and social-level policies (e.g. laws, licenses, contracts). Policies are collections of rules, with rules of a small number of types combined to permit sharing of, processing of, and delegation of control over assets.

In order to be able to give permissions regarding assets to sites and parties, all of these objects must be uniquely identified. This is done by an identifier of the form \texttt{<type>:<ns>:<name>}. Object types include \texttt{party}, \texttt{site}, \texttt{asset}, collections and categories of these (see below), and \texttt{result}, which is covered in Section \ref{sec:policy_evaluation}. Each participating party owns a namespace, which is a DNS (sub)domain registered to it that is used for the \texttt{<ns>} field. It is the responsibility of that party to ensure that all its objects are given a unique (within the namespace and organisation) \texttt{<name>} field. Each object is considered to be owned by the owner of the namespace it is in. This is strictly an administrative ownership which establishes the party with primary control over the asset within the Mahiru system, it does not necessarily correlate to legal ownership of any intellectual property related to the asset.

Mahiru rules are relations between assets, sites, parties and other objects. For each rule type, objects of particular types are referred to, one of which is the \emph{subject} of the rule. Rules may only be made by the owner of their subject. This is enforced by a cryptographic signature attached to the rule, which can be verified against the signing certificate of the party owning the namespace the subject is in, as registered in the Registry and countersigned by a CA. These signatures are created by the parties when they create the rules, and can be created off-line if security requirements warrant the extra effort. A Trusted Platform Module could provide for a less secure but more convenient alternative, depending on the threat model.

\subsection{Organising objects}
\label{sec:organising_objects}

The first two rules presented here provide means to group assets. Mahiru has two kinds of groupings, collections and categories. Collections and categories of assets are considered assets themselves, so that they can be nested. Assets may be placed in an asset collection by an \textbf{InAssetCollection}(\emph{asset}, collection) rule. The asset is the subject of the rule, and asset collections are identified by an identifier of type \texttt{asset\_collection}. Assets may be categorised by an \textbf{InAssetCategory}(asset, \emph{category}) rule, which has the category (of type \texttt{asset\_category}) as its subject.

Thus, asset collections and categories both group assets, but asset owners place their assets in collections, while category owners place assets in their categories. Permissions propagate through collections (i.e. if one has access to an asset collection, one has access to all assets in that collection, recursively) but not through categories. The distinction between collections and categories becomes relevant when the owner of the asset category or collection is not the same as the owner of the asset. Such cross-namespace rules are used to delegate authority, as described below.

Besides assets, also sites and parties can be categorised by corresponding rules. A site can be put into a site category by the owner of that category via an \textbf{InSiteCategory} rule, and a party can be put into a party category via an \textbf{InPartyCategory} rule. Unlike for assets, permissions propagate down through categories of sites and parties, so giving a particular site category access to an asset gives access to all sites in the category. The exact algorithm is explained below.

\subsection{Access and usage rights}
\label{sec:access_rights}

Mahiru features two kinds of permissions: access permission and usage permission. Access permission for an asset is given to a site, and permits the asset to be present in the Mahiru software on that site and, for software assets, for them to be run. This permission is given through a rule of the form \textbf{MayAccess}(\emph{asset}, site). Here, \emph{asset} is an asset or asset collection, while \emph{site} is a site or a site category, and the permission applies to any asset in the collection or any subcollections, and any site in the category or subcategories. Since assets can only be placed in collections by their owner, access permissions for a collection may only be given by its owner, and owners of site categories determine which sites are in them, it is clear that asset owners have ultimate control over the path from asset to site.

Usage permission for an asset is given to a party, using a rule of the form \textbf{MayUse}(\emph{asset}, party, conditions). Asset collections and party categories apply as above. This rule permits the named party to extract the asset from a Mahiru site and use it according to the given conditions (e.g. only for non-commercial use). Conditions may be expressed in plain text, or some formal description language such as the Open Digital Rights Language (ODRL) could be used. Note that by definition, the conditions cannot be enforced by Mahiru. Instead, they represent a non-binding request, an IP licence, and/or a reminder of a separately agreed contractual obligation.

Wildcards may be used for the non-subject objects in these rules, which makes it possible to declare assets to be publicly available.

\subsection{Rules for data processing}
\label{sec:processing_rules}

In Mahiru, data is processed using directed acyclic graph-style workflows, each step comprising the application of a compute asset (software) to one or more data assets (a data set or previous result). This produces results, which can in turn be inputs for another compute asset in the same workflow. Each step in the workflow must be executed by a site, which requires the presence of the step's inputs, the compute asset, and the step's outputs at that site. For this to be permitted, the site must have access rights (as described previously) to all of these. This poses a problem, because access rights for the outputs must be set before the workflow is executed\footnote{manually approving individual requests when needed works in principle, but does not scale}, at which time the outputs do not exist and cannot be referred to by a rule.

This is solved through the use of a \textbf{ResultOfDataIn}(\emph{data\_asset}, compute\_asset, output, collection) rule. With this rule, the owner of the specified data asset declares that if the asset is processed using the given compute asset then the asset produced for the named step output must be considered to be in the given collection. Access and usage permissions can then be given for this collection. The data asset may be a collection of assets instead, and for the compute asset an asset category may be given, with the rule applying recursively.

A second rule type, \textbf{ResultOfComputeIn} has the same attributes but has the compute asset as its subject, allowing its owner to give permissions as well. In this case, an asset category may be specified for the data asset, and an asset collection for the compute asset. Permission is needed from both the owner of the data asset and the owner of the compute asset for a compute step to be allowed to run.

Wildcards may be used here as well, so that for example permission to process a data asset $A$ in arbitrary ways on any site may be set by a combination of InAssetCollection($A$, $C$), ResultOfDataIn($C$, *, *, $C$), and MayAccess($C$, *).

\subsection{Delegation of authority}
\label{sec:delegation}

In some cases, asset owners may wish to delegate some or all of their authority regarding their assets to another party. For example, a data owner may have joined a consortium providing a pool of data assets to others, with access controlled by the consortium organisation, or they may want to delegate inspection of compute assets to be used to process their data to a trusted auditor. In Mahiru, this is done by creating rules which refer to (usually) collections and categories in different namespaces. In the first example, the consortium would set up a site, and allocate an asset collection identifier $C$ in their namespace. The consortium members would then add InAssetCollection($A$, $C$) for each asset $A$ they contribute to the consortium, thus enabling the consortium to give rights to $A$ via rules of the form MayAccess($C$, site) and MayUse($C$, party). For the second case, the auditor would set up a site and create a category $T$ of trusted software, then add an InAssetCategory(software, $T$) rule for each inspected and approved compute asset. Data owners would then refer to $T$ in their ResultOfDataIn rules to delegate the assessment of compute assets.

\subsection{Policy evaluation}
\label{sec:policy_evaluation}

When a workflow execution request or a data access request arrives at a site, the combined policies must be evaluated in order to determine whether to grant the request. For a data access request, this means verifying that the requester has access permissions for the requested asset. For a workflow execution request, the site must verify for each step assigned to it whether that step is allowed to run there, which entails verifying that it has access permissions to the step's inputs, the compute asset, and the step's outputs. To plan a workflow submitted by a local user, for each step the set of sites with permission to run that step is determined, after which possible plans can be enumerated by backtracking through the graph, or some optimal plan can be devised via dynamic programming.

From the above, it is clear that the central policy evaluation operation is to determine whether a particular site $s$ has access to a particular asset $a$. For \emph{primary assets}, which are data and compute assets that were put into the system directly, this is relatively straightforward. Define relations $M$ and $C_{oA}$ as

\begin{equation}
(s, a) \in M \; \textit{iff there exists a rule MayAccess(a, s)}
\end{equation}

\begin{equation}
\begin{array}{rll}
(a, c) \in C_{oA} & \textit{iff} & a = c, or \\
                  &              & \textit{a rule InAssetCollection(a, c) exists}
\end{array}
\end{equation}

and $C_{tS}$ analogously to $C_{oA}$ but for \textit{InSiteCategory} rules. We can then define the has-access relation $H$ as

\begin{multline}
(s, a) \in H \Leftrightarrow \exists (s', a') \\
    (s, s') \in C^+_{tS} \wedge (s', a') \in M \wedge (a, a') \in C^+_{oA}
\end{multline}

where superscript plus denotes the transitive closure. Informally, $s$ has access to $a$ if there is a path from $s$ directly or up through the site categories to a MayAccess rule referencing $a$ or an asset collection that directly or indirectly contains $a$.

\emph{Secondary assets} are assets derived from other assets, specifically intermediate and final results of workflows. Since these assets do not yet exist when the policies are set, there are no direct MayAccess rules referring to them. Instead, they are implicitly added to asset collections through the use of ResultOfDataIn and ResultOfComputeIn rules, and those rules must be combined with the definition of the workflow that produced the asset to determine which sites can access it. For each workflow step output, access permissions must be obtained from the owner of the compute asset used, the owner of each input, and the owners of assets used to produce each input, recursively.

\emph{Permissions} for result $r$ of a workflow step are represented by a permissions object $P(r)$, which is a set of sets of assets or asset collections. Given $P(r)$, we can define a second has-access relation $H'$ as

\begin{equation}
(s, r) \in H' \Leftrightarrow \forall C\!\in\!P(r) \:\: \exists c\!\in\!C \:\: (s, c)\!\in\!H
\end{equation}

In words, a site $s$ needs to have access to at least one asset or collection in each set in $P(r)$ to be able to access result $r$. For a compute asset $a$, which is always a primary asset, we set $P(a) = \{\{a\}\}$, so that $(s, a) \in H' \Leftrightarrow (s, a) \in H$. 

It remains to define how to propagate $P$ across workflow steps. This is done via the ResultOfComputeIn and ResultOfDataIn rules, for which we will define two more relations:

\begin{equation}
\begin{array}{rl}
(i, p, n, c') \in R_C & \textit{iff there exists a rule} \\
& \textit{ResultOfComputeIn}(i', p', n', c') \textit{ where} \\
& (i' = * \vee (i, i') \in C_{tA}^+) \; \wedge \\
& (p' = * \vee (p, p') \in C_{oA}^+) \; \wedge \\
& (n' = * \vee n' = n)
\end{array}
\end{equation}

\begin{equation}
\begin{array}{rl}
(i, p, n, c') \in R_D & \textit{iff there exists a rule} \\
& \textit{ResultOfDataIn}(i', p', n', c') \textit{ where} \\
& (i' = * \vee (i, i') \in C_{oA}^+) \; \wedge \\
& (p' = * \vee (p, p') \in C_{tA}^+) \; \wedge \\
& (n' = * \vee n' = n)
\end{array}
\end{equation}

These relations contain any combination of input, processing compute asset, output name and output collection for which a matching rule exists. Note that rules may specify wildcards ($*$) for the data asset, compute asset and output name. Also note that the use of categories and collections is reversed between the two rules because the subjects are reversed; in both cases a collection can be used for the subject of the rule, and a category for the other asset. The target $c'$ must be an asset collection.

Given a result $r$ produced by applying compute asset $p$ to inputs $I$ and taking its output $n$, we can compute $P(r)$ as 

\begin{equation}
\begin{array}{rl}
P_C(r) = \{ & \\
& \{c' \mid (c, p, n, c') \in R_C \wedge c \in C \} \\
& \quad \mid C \in P(i) \wedge i \in I\}
\end{array}
\end{equation}

\begin{equation}
\begin{array}{rl}
P_D(r) = \{ & \\
& \{c' \mid (c, p, n, c') \in R_D \wedge c \in C \} \\
& \quad \mid C \in P(i) \wedge i \in I\}
\end{array}
\end{equation}

\begin{equation}
P(r) = P_C \cup P_D
\end{equation}

Intuitively, we propagate each permission set in each $P(i)$ via matching ResultOfComputeIn and ResultOfDataIn rules to a set of output collections. Recall that access to at least one collection in each set in $P(r)$ is needed, so that each owner of an input and the owner of the compute asset gets control over the result. $P$ is computed recursively for the whole workflow, so that access and usage of the end result are only granted if all owners agree.

\section{Use cases}
\label{sec:use_cases}

With its ability to run arbitrary workflows in a distributed fashion, Mahiru allows for a variety of different execution patterns or archetypes.\cite{shakeri_2019} In this section, we demonstrate how various common patterns would be implemented in a Mahiru data exchange.

\begin{figure*}
\centering
\includegraphics[width=0.9\textwidth]{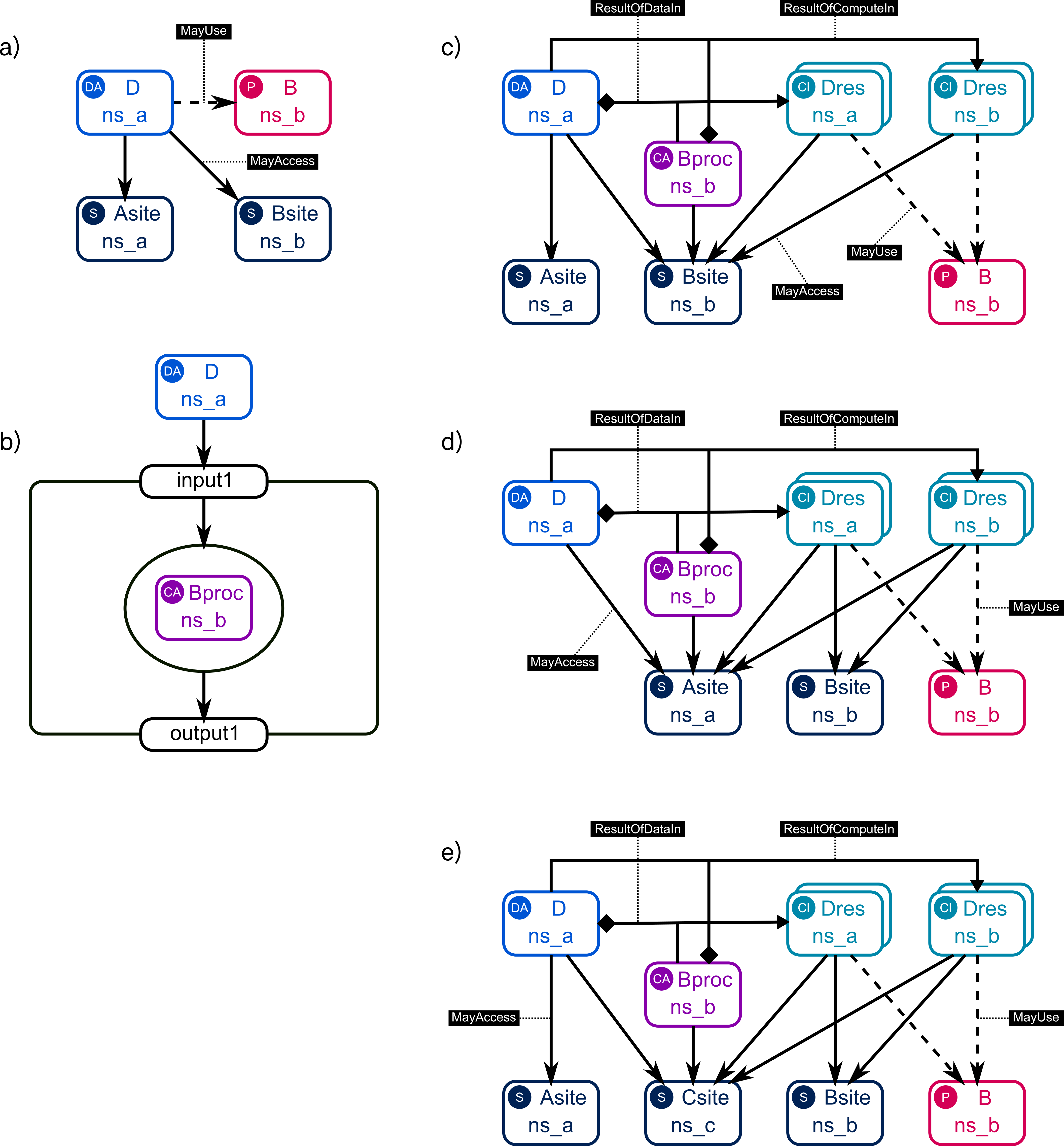}
\caption{Downloading and simple processing of data.
\enspace\textbf{a)} Policy rules needed for \textsf{B} to download and use dataset \textsf{D} from party \textsf{A}. Sites \textsf{Asite} (where the data originates) and \textsf{Bsite} (where it is copied to) are permitted to have a copy of \textsf{D} through MayAccess rules, and a MayUse rule permits party \textsf{B} to take it out of the Mahiru system.
\enspace\textbf{b)} A simple processing workflow comprising a single processing step using \textsf{B}'s compute asset \textsf{Bproc} applied to \textsf{A}'s dataset \textsf{D}.
\enspace\textbf{c)} Policy rules to allow running the one-step workflow at site \textsf{Bsite}. \textsf{Asite} and \textsf{Bsite} are again given access rights, allowing \textsf{D} to be copied to \textsf{Bsite}. Then \textsf{D} is allowed to be processed using \textsf{Bproc} by both \textsf{A} (via a ResultOfDataIn rule) and \textsf{B} (via a ResultOfComputeIn rule), with the result declared to be in the collections \textsf{ns\_a:Dres} and \textsf{ns\_b:Dres} respectively, to both of which \textsf{Bsite} needs access for the step to be run at \textsf{Bsite}. Finally, \textsf{B} needs to be permitted to use the result for it to be able to exit the system.
\enspace\textbf{d)} Permissions for running the same workflow but with the step executed at \textsf{Asite} (compute-to-data). Data, software and output are permitted to be at \textsf{Asite} through MayAccess rules, while the output is also permitted to go to \textsf{Bsite}, and party \textsf{B} is allowed to use it.
\enspace\textbf{e)} Policy for allowing a trusted-third-party scenario at \textsf{Csite}. \textsf{A} and \textsf{B} do not share their data and software with each other, but allow them to be copied to \textsf{Csite}, which can also have the results. Processing can then take place there, with the results once again permitted to go to \textsf{B} via \textsf{Bsite}.
}
\label{fig:one_step_workflow}
\end{figure*}

\begin{figure*}
\centering
\includegraphics[width=0.9\textwidth]{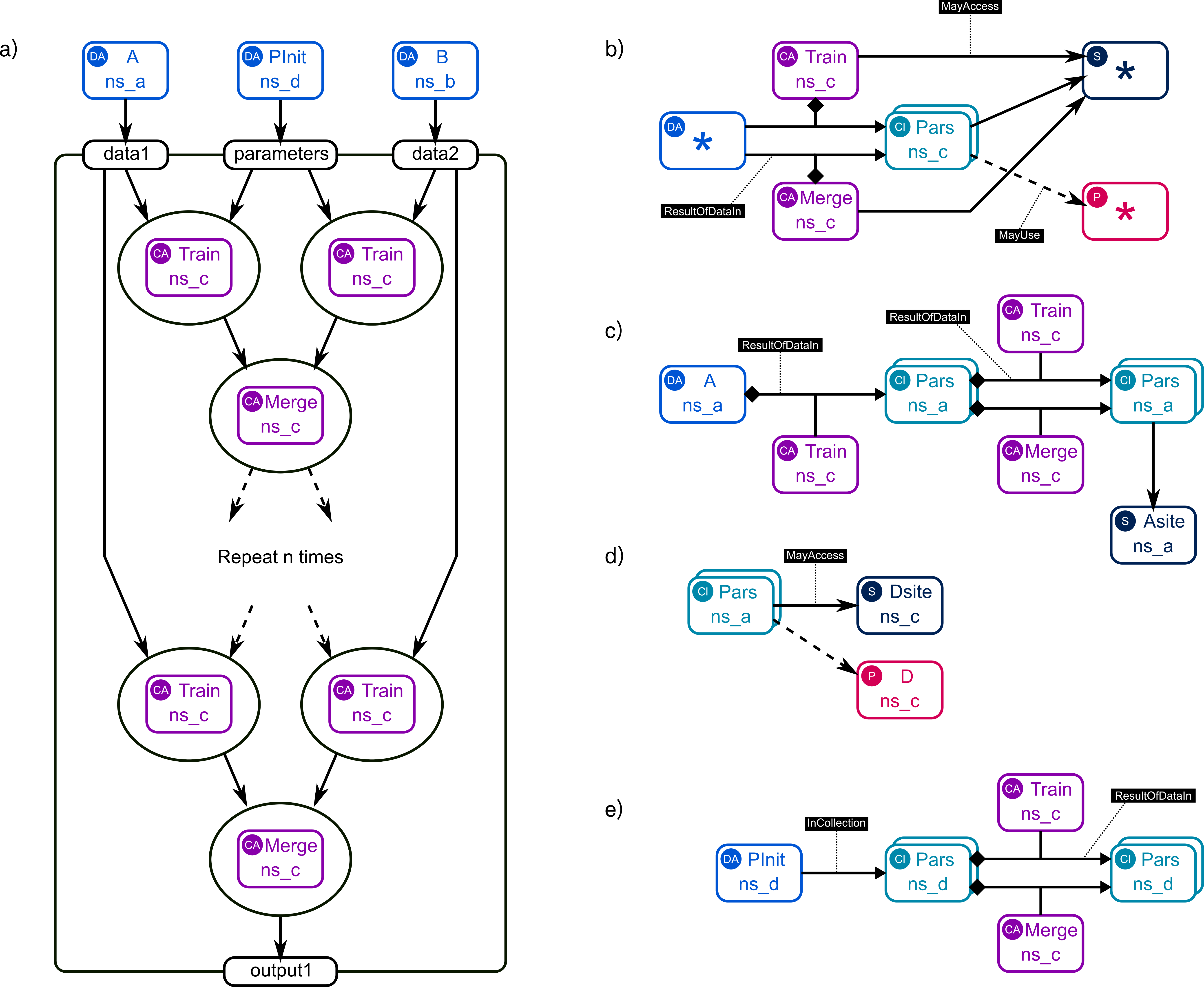}
\caption{Federated machine learning using Mahiru.
\textbf{a)} Workflow for horizontally partitioned federated learning. An initial set of model parameters \textsf{PInit} from a user at party \textsf{D} is updated based on input datasets \textsf{A} and \textsf{B} provided by parties \textsf{A} and \textsf{B} respectively, after which the updated parameters are combined and the process is repeated until convergence. The software (compute assets \textsf{Train} and \textsf{Merge}) is supplied by party \textsf{C}.
\enspace\textbf{b)} Permissions set by software supplier \textsf{C}. We assume here that the software is Open Source, and that any data may be processed with it and the results used by anyone. Different licensing models may be expressed through a policy that limits use to specific inputs or which only allows specific parties to use the results.
\enspace\textbf{c)} Policy set by data supplier \textsf{A} to allow local (at \textsf{Asite}) processing of its data using \textsf{C}'s software. Data asset \textsf{A} may be used as an input for the \textsf{Train} step, and the resulting updated parameters may be processed further repeatedly using \textsf{Train} and \textsf{Merge} steps. The other data supplier \textsf{B} sets identical policies.
\enspace\textbf{d)} Policy set by \textsf{A} to allow the results to be copied to \textsf{Dsite} and used by party \textsf{D}.
\enspace\textbf{e)} Policy set by \textsf{D} to allow the initial parameters to be used in the workflow.
}
\label{fig:federated_learning}
\end{figure*}

\subsection{Downloading data}
The simplest operation that can be performed by Mahiru is a data download. This is achieved by the data user submitting an empty workflow, which has one output connected to its one input and no processing steps. The desired data asset is set as the input, and the output will simply be the requested data. To be able to run this workflow, the requester and their site need MayUse and MayAccess permissions for the data asset (see Figure~\ref{fig:one_step_workflow}a).

\subsection{Local processing}
The previous case can be extended to do data processing by adding a step to the workflow that runs a compute asset, possibly created by the data user (Figure~\ref{fig:one_step_workflow}b). If the user's site can access the compute asset that is used in the step as well as the step's output, and the user may use the result, then this workflow can be run by downloading the data asset to the user's site and then running the step there. Note that the data owner needs to give permission for both the transfer of the input data and for giving the output to the data user (Figure~\ref{fig:one_step_workflow}c).

\subsection{Compute-to-data}
Giving the user permissions to obtain a copy of the data requires a large amount of trust, because the data owner has no control over the data user's system. Alternatively, the exact same workflow could be run with the step running at the site of the data owner. This way, no access permissions to the data are needed for the user's site. However, the data owner's site must now be able to access the compute asset as well as its output, and the data user's site still needs to be able to access the output (Figure~\ref{fig:one_step_workflow}d). If these permissions are available, then the workflow can be executed by sending the compute asset to the data owner's site, executing the step there, then downloading the result to the user's site. Note that if the roles of the software and the data are swapped, i.e. the software owner not allowing the software to go to another site and the data being sent to it from the user's site, processed, and the result returned to the user, then we obtain a Software-as-a-Service archetype (not shown).

\subsection{Trusted third parties}
In some cases, the data owner does not trust the data user with a copy of the data, and the data user does not trust the data owner with a copy of their software. In this case, if a third party can be found that is trusted by both the data owner and the data user, then permissions can be set accordingly and the processing can be done by this trusted third party (Figure~\ref{fig:one_step_workflow}e). Note that there may be other reasons for using a trusted third party, such as the availability of large compute resources\cite{Scheerman_ODISSEI}. If the owner of the data and the software is the same party in such a case, then this setup becomes an Infrastructure-as-a-Service model (not shown).

\subsection{Federated machine learning}
Federated machine learning is a somewhat more complicated case, because the algorithm involves many processing steps. For horizontally partitioned data, a model can be trained by repeatedly sending it to the data sites, updating its coefficients there using the local data, then sending back the new coefficients and combining them with the ones obtained from the other data sites. This can be expressed in a Mahiru workflow which alternates a set of parallel training steps with combining steps (Figure~\ref{fig:federated_learning}a). This workflow involves four parties: data providers A and B, software provider C, and data user D. C's policies govern the software, where it can go and how it can be used, with a very liberal policy shown in Figure~\ref{fig:federated_learning}b). The data providers allow C's software to be used with their data (Figure~\ref{fig:federated_learning}c), and they allow D to have the results (Figure~\ref{fig:federated_learning}d). Finally, D needs to allow its inputs to be used (Figure~\ref{fig:federated_learning}e). If the data cannot leave their owners' sites, then the training steps have to run at the data sites in a compute-to-data fashion, while the combining steps can go anywhere the data owners allow the coefficients to go. If additional permissions are available then other ways of executing the workflow may be possible too. For example, if one of the data owners has given a trusted third party's site access to the data, then the training steps for that data set can be run at this site as well, and possibly more efficiently. The policy evaluation algorithm will automatically take this into account, and produce corresponding execution plans.

\section{Implementation}
\label{sec:implementation}

To demonstrate the feasibility of our design, we have created a proof-of-concept implementation, which is available online as Open Source software\cite{Mahiru_software}. This implementation lacks some of the features needed for production use (such as persistent storage, and user interfaces), but does demonstrate the policy mechanism (including replication), data exchange, and distributed workflow execution.

A Mahiru data exchange consists of a registry and a collection of sites. The registry is a database that can be accessed through the simple replication protocol described in section~\ref{sec:registry}. The amount of stored data is tiny and update rates of the database are likely to be very low, as a result of which most requests for updates from the sites will be empty and can be generated quickly. Also, in most cases propagation delays of many hours will be acceptable, so that even for a large exchange with many sites the load on the system is low. A simple database server therefore suffices.

\begin{figure}
\centering
\includegraphics[width=\columnwidth]{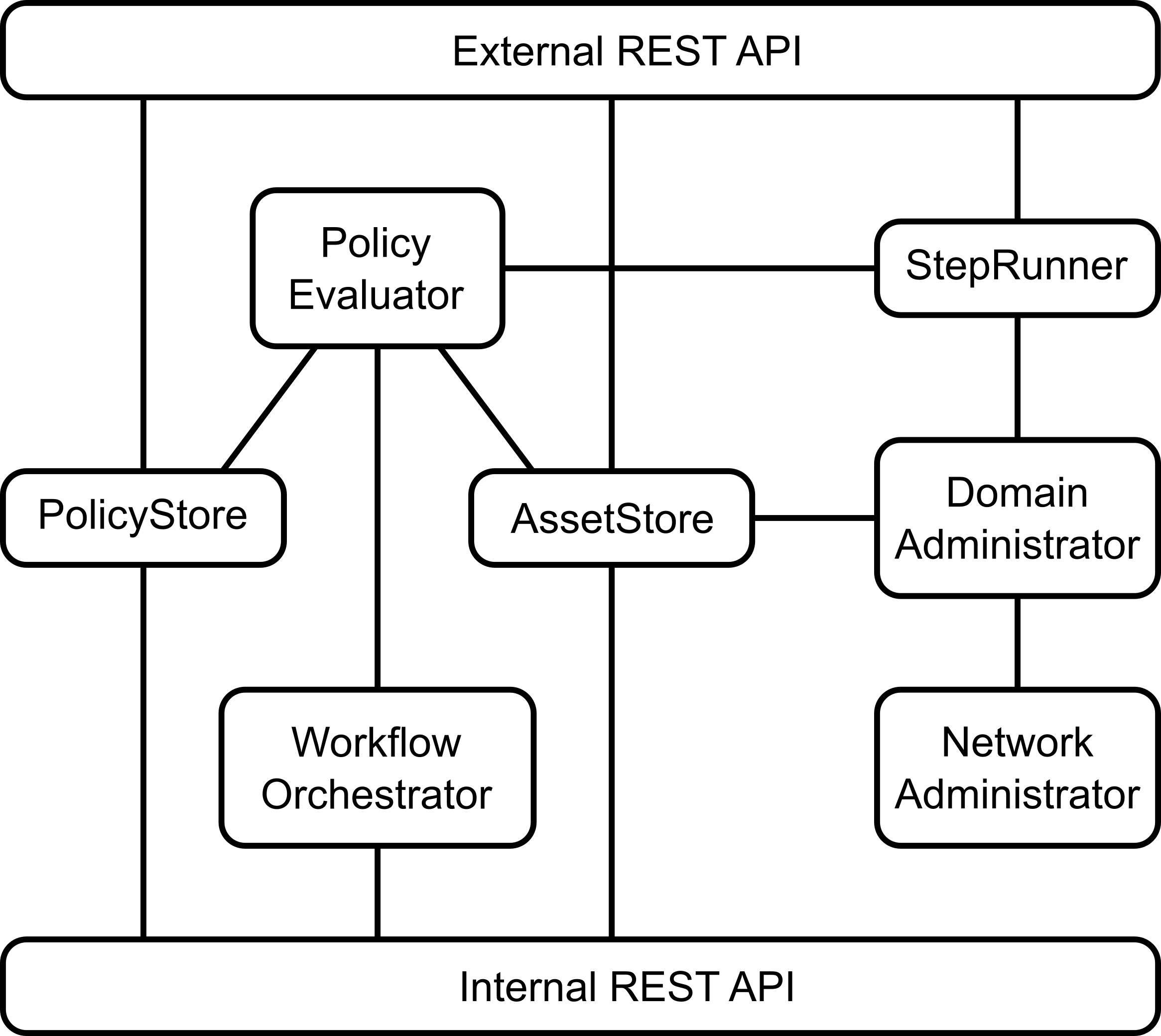}
\caption{Mahiru site software architecture. The PolicyStore, PolicyEvaluator and AssetStore implement data exchange, the WorkflowOrchestrator orchestrates distributed workflow execution, and the StepRunner, DomainAdministrator and NetworkAdministrator execute workflow steps. See Section~\ref{sec:implementation} for details.}
\label{fig:site_architecture}
\end{figure}

The sites perform all the work of sharing and processing data, and are somewhat more complicated. Figure~\ref{fig:site_architecture} shows the main components of a site in our reference implementation. The software is envisioned to be deployed in a DMZ, exposing an External REST API to the Internet, and an Internal REST API to the Intranet.

The PolicyStore component contains a database for the local rules and a replication server accessible through the External REST API for other sites to replicate the local policies. It also manages a set of replicas of other sites' policy databases. An internal API endpoint accessible through the Internal REST API can be used for setting the local policies. Attached to the PolicyStore is the PolicyEvaluator, which implements the algorithms described in section~\ref{sec:policy_evaluation}, and serves as an authorisation server to the AssetStore, WorkflowOrchestrator, and StepRunner.

The AssetStore stores data and compute assets as well as their metadata. It exposes an API endpoint for managing the assets on the Internal REST API, and an API endpoint for retrieving the assets on the External REST API. It uses the PolicyEvaluator to authorise any external requests. Both data and compute assets are Docker container images.

The WorkflowOrchestrator can be accessed through the Internal REST API, taking workflow execution requests from a user application running on the Intranet. For each request, it creates an execution plan as described in section~\ref{sec:policy_evaluation}, calling on the PolicyEvaluator to evaluate the combined policies. Once a plan has been made, it sends step execution requests to the required sites' External REST APIs to orchestrate the execution.

These step execution requests are routed to the StepRunner, which first uses the PolicyEvaluator to verify their legality, and then uses the DomainAdministrator to locally orchestrate the containers needed to execute the step. The DomainAdministrator first creates a local network containing a container for each input asset's image as well as a container for each output, and then runs the compute asset container inside the same bridge network. The data asset containers run an HTTP server serving the data, and the compute asset container uses this to retrieve its inputs. Outputs are written to a WebDAV-enabled HTTP server in the output containers, which are then saved to images for use by the next step. The prototype does all this using the local Docker service.

In many cases, a compute asset may only need a small fraction of the data available in a particular data asset. In this case, downloading the entire image is wasteful, especially if the data set is large. The prototype therefore allows accessing an input data asset remotely. To do this, the StepRunner sends a connection request rather than a download request to the site holding the input asset, and that site's AssetStore has its DomainAdministrator create a container for it. Through their NetworkAdministrator components, the two sites then set up a WireGuard VPN connection between it and the compute asset container on the executing site, after which execution proceeds as described before.

A production version of the system will need to service many more requests, and therefore a larger number of containers are needed. Kubernetes and Docker Swarm are commonly used solutions for managing large numbers of containers across many physical machines. Kubernetes and Swarm employ overlay networks technologies to connect containers together and control their connectivity via network policies. This allows connecting containers belonging to the same step execution request while isolating them from containers belonging to other requests \cite{LCN2020}. Like Kubernetes and Docker Swarm, these overlay networks are centrally controlled and not designed to operate across independently administrated domains. As such, they cannot be used to provide the cross-domain connections to remote data assets. Instead, WireGuard VPN tunnels can be used as in the current prototype, at the cost of having to manage the two solutions to make them work together.

As an alternative approach, we have investigated the use of P4 programmable switches to facilitate both the local and the remote connections\cite{CITS2022}. This could potentially facilitate both within-domain and cross-domain connectivity with a single solution, and it enables the use of programmable network hardware like SmartNICs or DPUs to offload packet processing and improve performance. Furthermore, P4 programs can be used to audit and constrain the network activity of the containers to provide isolation and improve system trust.

\begin{figure}[t]
\centering
\includegraphics[width=\columnwidth]{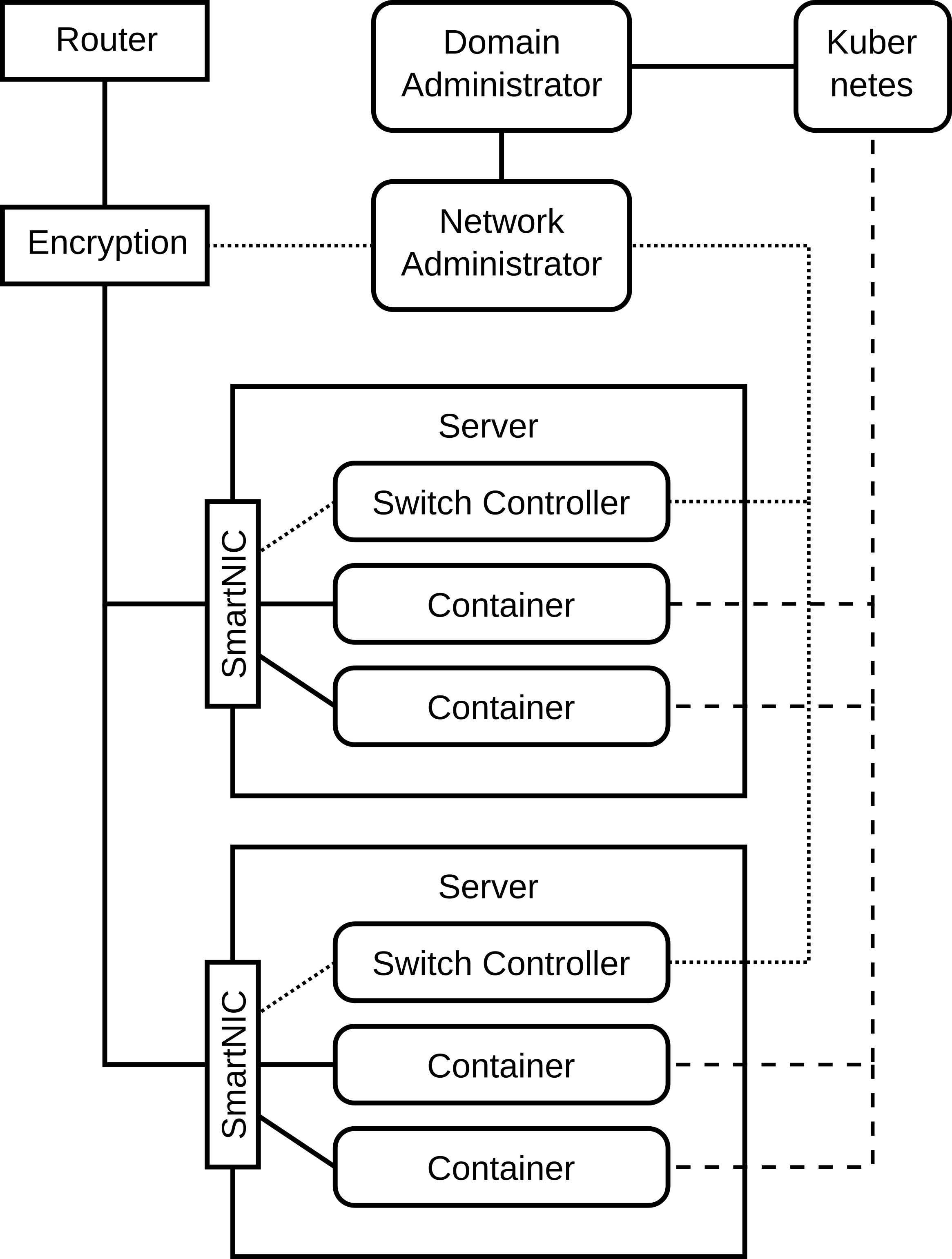}
\caption{Proposed design for a scalable container execution backend. Containers are run on multiple servers equipped with SmartNICs, which are programmed to connect the containers to each other within and between servers, and to containers running at other sites via a separate encryption device and router. Mahiru's Domain Administrator controls the containers via Kubernetes, and its Network Administrator controls the networking hardware. A Switch Controller component on each server programs the local SmartNIC on behalf of the Network Administrator.}
\label{fig:network_architecture}
\end{figure}

Figure~\ref{fig:network_architecture} shows how a more scalable implementation orchestrates containers and network connections. The DomainAdministrator starts containers by contacting Kubernetes, which in turn starts the required containers on one or more of the servers it controls, and returns their locations. The DomainAdministrator then sends this information to the NetworkAdministrator, which sets up connections between the containers. If two containers are on the same server, then the SmartNIC in that server can be used to connect them directly. For containers on different servers, the corresponding SmartNICs need to collaborate to make the connection. If a container needs to connect to a remote container, its local SmartNIC is programmed to set up a connection through the Encryption service, the Router and the Internet to the other domain. In this latter case information is needed on how to connect to the other domain, which is obtained by Mahiru through its site-to-site API connection and passed to the DomainAdministrator as well. For details we refer to Shakeri et al.\cite{Shakeri2022}. To support auditing, the NetworkAdministrator would program the SmartNICs to inspect traffic and log activity, and a new component would be added to retrieve these logs, analyse them and provide the results to the system administrator \cite{CITS2022}.

\section{Discussion and future work}
\label{sec:discussion}

In this paper, we have introduced an architecture for a federated data exchange and a policy mechanism supporting it. The architecture is based on standard technologies, including REST-based web services and public-key cryptography. It also uses containers and WireGuard, which is gaining acceptance as the new standard for secure network tunnels. A proof-of-concept implementation demonstrates the feasibility of this design, although scalability remains to be fully demonstrated.

\subsection{Policies}

Regarding basic access to assets the permission system resembles a simple discretionary access control system as found in file systems and file servers, with groups of objects and groups of subjects that can access them. Propagation of access and usage rights to the members of a collection enables delegation of control, which could also be achieved with e.g. POSIX file permissions. An extension is needed and provided to be able to authorise data processing however, with a minimal increase in complexity. All policy rules are permissions, which simplifies the system and reduces the risk of unintended consequences when policies are combined across organisational boundaries, but it does mean that prohibitions and duties cannot be expressed.

Highly complex legislation, standards and contracts often govern data exchange and must be dealt with. More complex access control systems like XACML\cite{XACML}, Organization Based Acces Control\cite{OrBAC} or eFlint \cite{eFlint} allow much more complex policies, at the cost of increased implementation complexity. They also make the policies more difficult to understand, and unintended consequences of combining policies made by different administrators harder to avoid. Some work has been done on combining XACML policies, in a simplified case in the Web Services Policy Language (WSPL)\cite{Anderson2004} and also for more complex scenarios\cite{Mazzoleni2008}. One interesting possibility would be to use one of the more complex languages to express a very detailed policy, and then to derive a Mahiru policy from this. We are currently exploring this possibility with the developers of eFlint.

One other interesting difference between Mahiru's policy mechanism and XACML is the use of an eventually-consistent replication mechanism to exchange policies between organisations. XACML assumes that policies are evaluated by a server named a Policy Decision Point (PDP) and enforced by a Policy Enforcement Point (PEP). Mahiru sites internally use a similar design, with the PolicyEvaluator as the PDP and the AssetStore and StepRunner as PEPs. Doing this across sites to facilitate delegation of authority and transfer of assets to other sites would result in the load on the PDP increasing with the number of workflows being executed throughout the system (workflow planning in particular requires many policy decisions as alternative possibilities are evaluated). Having effectively a cache at each site means that each request to the replication server covers many workflows.

\subsection{Workflows}

The workflow format and execution engine implemented in the prototype are rather simple, and serve mainly to demonstrate the federated execution mechanism. For a production-ready system, several extensions are needed, and more can be made to widen the applicability of the system. First, workflow signing as described above is not yet fully implemented: the certificates are there but the signatures not yet. We plan to fix this in the near future. Error handling and progress monitoring, while not as important for a prototype, are certainly needed for a production system, as is accounting, which will also entail a small extension to the policy mechanism to enable owners of compute resources to give permission for parties to run workflows there. For increased security, another rule type can be added that lets owners of compute resources control which software assets they are willing to run.

Second, workflow inputs need to be added. Currently, workflows process only assets, which are static data sets managed by a system administrator. However, practical workflow runs often have some data that is specific to that particular workflow run and created by the user, for example configuration data for an execution step or a processing script in a compute-to-data scenario. Entering these as assets in impractical. Instead, it needs to be possible for the user to submit these ephemeral inputs with the workflow job itself, with the user giving implicit permission to process them according to the submitted workflow.

Third, assets may change over time as new data are collected or software is updated. Giving assets a version number in addition to their ID makes it possible to track these changes for the purpose of provenance. Policies might then be extended with version constraints, although this would increase complexity.

Secure multiparty computation can potentially be supported by Mahiru by extending the policy mechanism, the asset types and the workflow format to support shares of assets. Such a design must ensure that enough shares to reconstruct the asset can only end up on sites that can access that asset, ideally also between multiple workflow executions.

Adding loops to the workflow format would reduce the size of repetitive workflows such as those used in machine learning tasks. These would work much like loops in a programming language, and allow the loop body to be planned once, rather than repeatedly. If containers for workflow steps within the loop body are reused between steps, then the overhead of creating and destroying them is avoided, and by putting FIFO-queues in the data asset containers for the intermediate results in between steps and using remote access, a distributed application emerges that is still governed by the Mahiru policies. However, this does transfer some of the responsibility of executing the workflow from the Mahiru middleware to the assets, which requires increased trust and/or better monitoring of execution.

Loops could also facilitate stream processing, for example in an Internet-of-Things context, with each loop iteration processing a new item from an input stream and producing an item for an output stream. Beyond this, even distributed coupled simulations seem possible in which different parties collaboratively simulate a system (e.g. an energy grid) without sharing their simulation codes (which likely contain proprietary information about their network), as governed by policies enforced by their own system.

\section{Conclusion}
\label{sec:conclusion}

Data sharing is a complex problem. Organisations have many kinds of data, for which different kinds of sharing are appropriate. Maintaining multiple systems to support this is expensive and potentially failure-prone. In this paper, we introduced Mahiru, a design for a flexible data sharing architecture that can support many different data sharing methods. Mahiru is based on a federated architecture, in which workflows can be executed in a distributed fashion while adhering to policies set by owners of the data and software used. The policy mechanism is simple yet powerful, which makes it easier for users to set policies and understand their implications. Use of a replication algorithm in key places ensures scalability of the system. As a result, Mahiru can support sharing of public as well as sensitive data, a variety of processing models including compute-to-data, software-as-a-service, trusted-third-party, infrastructure-as-a-service, and federated learning, all based on a single design with a single policy mechanism.

\section*{Acknowledgements}

This research was funded by the Netherlands eScience Center and the Netherlands Organisation for Scientific Research under project 2717G18 SECCONNETSmart.

 

 \bibliographystyle{elsarticle-num} 
 \bibliography{cas-refs}





\end{document}